\documentclass[preprint,aps,floats]{revtex4}
\usepackage{epsfig}

\usepackage{graphicx}
\usepackage{dcolumn}
\usepackage{bm}
\def\lsim{\mathrel {\vcenter {\baselineskip 0pt \kern 0pt
    \hbox{$&lt;$} \kern 0pt \hbox{$\sim$} }}}
\def\gsim{\mathrel {\vcenter {\baselineskip 0pt \kern 0pt
    \hbox{$&gt;$} \kern 0pt \hbox{$\sim$} }}}
\def\beq{\begin{equation}}
\def\eeq{\end{equation}}
\def\bea{\begin{eqnarray}}
\def\eea{\end{eqnarray}}

\begin{document}

\title{Theta terms and asymptotic behavior of gauge potentials in (3+1) dimensions}

\author{Stephen D.H. Hsu}
\affiliation{Academia Sinica, Taiwan and Institute of Theoretical Science, University of Oregon}

\date{\today}


\begin{abstract}
\noindent We describe paths in the configuration space of (3+1) dimensional QED whose relative quantum phase (or relative phase in the functional integral) depends on the value of the theta angle. The final configurations on the two paths are related by a gauge transformation but differ in magnetic helicity or Chern-Simons number. Such configurations must exhibit gauge potentials that fall off no faster than 1/r in some region of finite solid angle, although they need not have net magnetic charge (i.e., are not magnetic monopoles). The relative phase is proportional to theta times the difference in Chern-Simons number. We briefly discuss some possible implications for QCD and the strong CP problem.
\bigskip

\noindent Revised version 2020: see Note Added after Conclusions.
\end{abstract}


\maketitle

\section{Physical consequences of Abelian theta terms?}

In an earlier paper \cite{theta} we proposed an interference experiment with outcomes sensitive to the value of the theta angle in QED. As depicted in figure 1, a combination of mirrors and beam splitters produces a superposition of two photonic states, which are then recombined to produce an interference pattern. The relative phase between the two states is equal to theta times the integral over $1/4 \, F_{\mu \nu} \tilde{F}^{\mu \nu} =  E \cdot B$, which is meant to be non-zero for the upper path (on which the photon packet passes through a background field, or encounters another photon packet) and zero for the lower path. This result can be deduced in two ways, either by considering a functional integral, or by solving the functional Schrodinger equation in QED (see, e.g., \cite{Treiman:1986ep}). In the former case, we use
\begin{equation}
\langle A_f \vert U_\theta \vert A_i \rangle =  \int_{A_i}^{A_f} DA~  \exp \left( - i \int d^4x ~  {1 \over 4} F^2  +   \theta  E \cdot B  \right)  ~~~,
\end{equation}
where $U_\theta$ is the time evolution operator in the presence of a theta term. In the latter case, we note that the theta term does not alter the QED Hamiltonian, but does shift the canonical momentum by $\theta B(x)$, where $B(x)$ is the magnetic field. Solutions to the Schrodinger equation in the presence of $\theta$ can be related to $\theta = 0$ solutions as follows
\begin{equation}
\label{Psiphase}
\Psi_\theta [A] ~=~ \exp \left(   i \theta \int d^3x \, {A \cdot B \over 2} \right) \Psi_{\theta = 0} [A] ~~.
\end{equation}
We recall that the current $K^\mu = 1/4 ~ \epsilon^{\mu \alpha \beta \gamma} F_{\alpha \beta} A_\gamma$ has divergence $\partial_\mu K^\mu = 1/4 \, F \tilde{F} = - E \cdot B$, so the spacetime integral over $E \cdot B$ can be re-expressed in terms the difference between spatial integrals at early and late times:
\begin{equation}
\label{EB}
- \int d^4x ~E \cdot B = \int d^3x~ \bigg[ K^0 (A_f) -  K^0 (A_i) \bigg] ~ \equiv ~ 8 \pi^2 \bigg[ H(A_f) - H(A_i) \bigg].
\end{equation}
Note we work in Minkowski space and restrict ourselves to configurations where the integral of $\vec{K}$ at spatial infinity ($\vert \vec{x} \vert \rightarrow \infty$) can be neglected. $K^0 = \frac{1}{2} \, A \cdot B$ is the same density that appears in the exponent in (\ref{Psiphase}). Its integral over space is  $8 \pi^2 H(A)$, where $H(A)$ is the magnetic helicity (or Abelian Chern-Simons number) \cite{MH} of $A$, which counts the linking number of magnetic field lines.

\begin{figure}[ht]
\includegraphics[width=10cm]{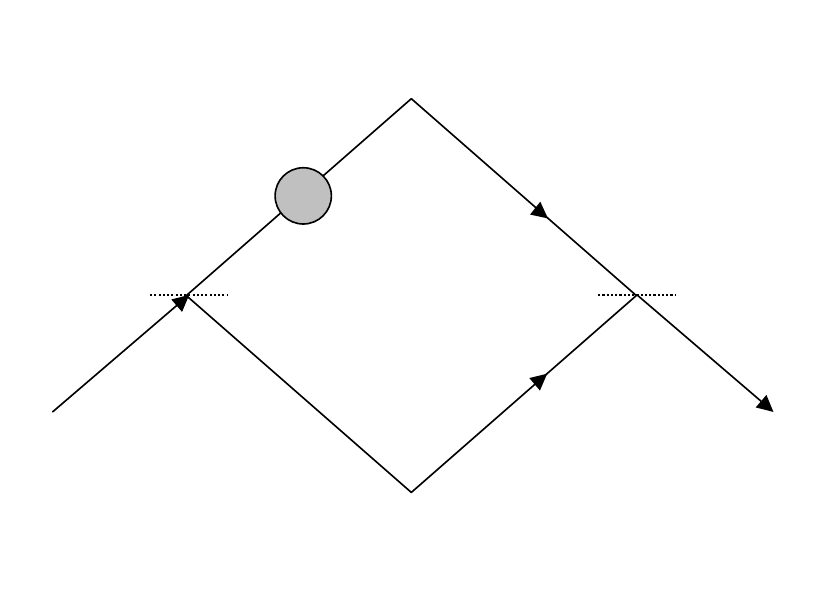}
\caption{A superposition of two identical wave packets of light are sent along upper and lower paths of the same length. The upper packet is exposed to a background electromagnetic field depicted by the shaded circle. This background field is chosen so that $E \cdot B$ is non-zero in the interaction region. Could interference between the recombined packets depend on the parameter $\theta$?}
\label{figure1}
\end{figure}

Contrary to the author's expectations in the earlier paper \cite{theta}, it is difficult to obtain a non-zero integral of $E \cdot B$ in a laboratory setting. Using (\ref{EB}), we see that it is necessary that the magnetic helicity be different at early and late times. Typical configurations produced by, e.g., lasers or atomic transitions, such as individual Fock states, have zero magnetic helicity.

We can reach a similar conclusion via a slightly different argument: consider any configuration that can be expressed in terms of a sum of waveforms: $A \sim \sum_i f_i( \hat{k}_i \cdot x - t ) g_i (x_\perp)$, where $\hat{k}_i$ is the direction of propagation, $x_\perp$ are the perpendicular coordinates, and the $f_i$ and $g_i$ are assumed to be localized. Since photons do not interact classically, these waveforms propagate undisturbed and become widely separated at early and late times. The magnetic helicity, which is the integral over all space of $~\epsilon^{ijk} \, A_i \partial_j A_k~$, decomposes into integrals over widely separated regions, each of which yields the same result at early and late times due to the form of the functions $f_i$ and $g_i$. The same is true if we consider wavepackets which pass through a static background field. Thus, it seems quite challenging to produce, in a real experiment, an effect which is sensitive to theta.

\bigskip

In what follows we describe specific gauge configurations whose quantum interference does depend on theta. These configurations are necessarily non-perturbative, with potentials that fall off only as $1/r$ in some non-zero solid angle. They are not ``realistic'' configurations that can be produced in the laboratory, and they presumably only lead to exponentially small effects in QED. Nevertheless their existence is evidence that the theta angle does have physical consequences.

\section{Asymptotic behavior of gauge fields in Minkowski space}

In the usual analysis of gauge theories potentials are required to fall off faster than $1/r$. However, there does not seem to be any strong justification for this other than convenience -- i.e., such configurations are amenable to topological classification. We relax this condition, and investigate configurations with the following properties.

\begin{enumerate}
\item the magnetic field $B(x)$ falls off as $1/r^2$ or faster at large $r$

\item the divergence of $B$ is zero everywhere (no monopoles)


\item they can be smoothly deformed into nearby configurations with faster (e.g., $A \sim r^{-(1+\delta)}$) fall off, and also to the vacuum $A = 0$, without passing through configurations with infinite energy.
\end{enumerate}

\bigskip

In Euclidean space configurations with $A \sim 1/r$ have infinite action, since the contribution to the action from $\int d^4x_{\rm E} ~E_i^2$ diverges. However, there does not seem to be any good reason to exclude such configurations in Minkowski space \cite{Mink}. Indeed, exact solutions with $1/r$ behavior are known in $SU(2)$ gauge theory \cite{LS,GH}.

\section{Jackiw-Pi configurations}

We make use of $d=3$ configurations constructed by Jackiw and Pi \cite{JP} by projecting SU(2) vacuum gauge fields (given by spatially varying group elements $U(r) \in$ SU(2)) onto a fixed ($\equiv$ electromagnetic) direction in isospin space. Explicit expressions for the vector potential and magnetic field in spherical coordinates are given below, with $f(r)$ an arbitrary scalar function. (The prime denotes differentiation with respect to $r$.)
\begin{eqnarray}
\label{JP1}
a_r &=& (\cos \theta) f'  \label{eq:2.23a}  \nonumber \\[1ex]
a_\theta &=& -(\sin \theta) \frac{1}{r} \sin f
\label{eq:2.23b} \nonumber \\[1ex] 
a_\phi &=& -(\sin \theta)  \frac{1}{r} (1-\cos f) 
\label{eq:2.23c} 
\end{eqnarray}

\begin{eqnarray}
b_r &=&-2 (\cos \theta) \frac{1}{r^2}  (1-\cos f)
\label{eq:2.24a} \nonumber \\[1ex] 
b_\theta &=&(\sin \theta)  \frac{f'}{r}  \sin f
\label{eq:2.24b} \nonumber \\[1ex] 
b_\phi  &=&(\sin \theta) \frac{f'}{r}  (1-\cos f)
\label{eq:2.24c} 
\end{eqnarray}
The magnetic helicity of a particular configuration is given by 
\beq
H= - \frac{1}{2\pi} \int^\infty_0  dr \, \frac{d}{dr}
(f-\sin f) = -\frac{1}{2 \pi}  (f-\sin f) \Big|_{r=\infty}
\label{eq:2.22}
\eeq
where we have taken $f(0)$ to vanish. When
$\sin f(\infty)$ is nonvanishing the magnetic helicity $H(A)$ can be an
irrational/transcendental number.  When $f(\infty)$ is an
even integer multiple of $\pi$,  $\sin f(\infty)$ vanishes and the helicity is an
integer.  An odd integer multiple of $\pi$ for $f(\infty)$
leads to vanishing $\sin f(\infty)$, and a half-integer value for $H(A)$. 
We will make use of configurations for which $f(\infty)$ is not an integer multiple of $\pi$. Such configurations have gauge potentials which fall off as $1/r$ and magnetic fields with $1/r^2$ behavior.

\section{A loop in configuration space}

Consider two configurations which differ by a gauge transformation: $A'_f (x) = A^\Omega_f (x)$. The magnetic helicities of the two differ by
\beq
\label{Hg}
\delta H = H(A'_f) - H(A_f) = \frac{1}{2} \int d^3x ~ \partial_i \Omega B^i = \frac{1}{2} \int d\Sigma_i ~\Omega B^i ~~,
\eeq
where $\Sigma$ is the surface at spatial infinity. Thus $\delta H$ can be non-zero when $B_i$ falls off no faster than $1/r^2$ in some region of finite solid angle. For such configurations $H(A)$ is not gauge invariant for appropriately chosen transformations $\Omega (x)$ with support at infinity.

In \cite{JP} it was shown explicitly that configurations given in (\ref{JP1}) with  $f( \infty ) / \pi$ non-integer 
have magnetic helicities which are not gauge invariant. For example, if one transforms to the Coulomb gauge, so that potentials in (\ref{JP1}) are transverse, the magnetic helicity is modified as follows
\beq
H (A^T) = H(A) + \frac{1}{6 \pi} (1 - \cos f) \sin f  \Big|_{r = \infty}~~.
\eeq
As expected from equation (\ref{Hg}), $H(A)$ is not necessarily gauge invariant when the magnetic fields fall off only as $1/r^2$, which in this case corresponds to $(1 - \cos f) \sin f  \Big|_{r = \infty}$ non-zero.

\bigskip

Now consider two paths in configuration space $A_1 (t,x)$ and $A_2 (t,x)$, taken to be equal at $t=0$: $A_1 (0,x) = A_2 (0,x) = 0$. Here $t$ is simply a path parameter and need not be the time coordinate. Suppose that $A_1 (t \rightarrow \infty, x) = A_f (x)$ and  $A_2 (t \rightarrow \infty, x) = A'_f (x) = A^\Omega_f (x)$. Joining the two paths together, we obtain a closed loop in configuration space, modulo a gauge transformation at $t = \infty$. (See figure 2.)

One example of such a pair is 
\bea
A_1 (t,x) &=& g(t) ~A_{JP} (x) \nonumber \\
A_2 (t,x) &=& g(t) ~A_{JP}^\Omega (x) ~~,
\eea
where $A_{JP} (x)$ is one of the Jackiw-Pi configurations of (\ref{JP1}), with non-integer  $f( \infty ) / \pi$, and the function $g(t)$ satisfies:
$g(0) = 0$, $g(t = \infty) = 1$. 

Note that at intermediate times $0 < t < \infty$ the two configurations $A_1 (t,x)$ and $A_2 (t,x)$ are not related by a gauge transformation, even though $A_2 (t = \infty,x) = A^\Omega_1 (t = \infty,x)$. For example, $E_1 (t,x) \sim \partial_t A_1 (t,x) = g'(t) \, A_{JP} (x)$, whereas $E_2 (t,x) \sim g'(t) \, A_{JP}^\Omega (x)$, so at intermediate times the $E$ fields of the two interpolations are not identical. We take $g'(t)$ to approach zero at early and late times. If the region where $g'(t)$ is non-zero has timelike extent $T$, then (assuming, e.g., $g(t)$ linear in $t$) the energy density at large $r$ from the electric fields is $\sim (rT)^{-2}$. If the timelike extent is taken to infinity appropriately (e.g., $T \sim R$, where $R$ is the spacelike extent) then all of the interpolating configurations will have finite energy.

It is easy to check that there is no flow of topological charge through the surface at spatial infinity: $0 < t < \infty$ and $r = | \vec{x} | \rightarrow \infty$. The Chern-Simons current flowing through this surface is given by
\beq
\int_0^\infty dt ~ r^2 d \Omega ~ K^{r} = \int_0^\infty dt ~r^2 d \Omega  ~\left[ \epsilon^{r \theta t \phi} A_\theta \partial_t A_\phi ~+\epsilon^{r \phi t \theta} A_\phi \partial_t A_\theta \right]
\eeq
Since the temporal component $A_0 = 0$ these are the only terms that contribute to $K^r$; the first and third indices of the epsilon tensor are fixed to be $r$ and $t$. Note that $\partial_t$ only acts on $g(t)$, so the form of the integrand above is (the lower case $a (x)$ denote Jackiw-Pi configurations $A_{JP}$):
\beq
g(t) a_\theta (x) \,  g'(t) a_\phi (x)  ~ - ~ g(t) a_\phi (x) \, g'(t) a_\theta (x)  ~=~ 0~.
\eeq
Thus, no net current flows through the timelike surface at $|\vec{x}| = \infty$. 

\begin{figure}[ht]
\includegraphics[width=10cm]{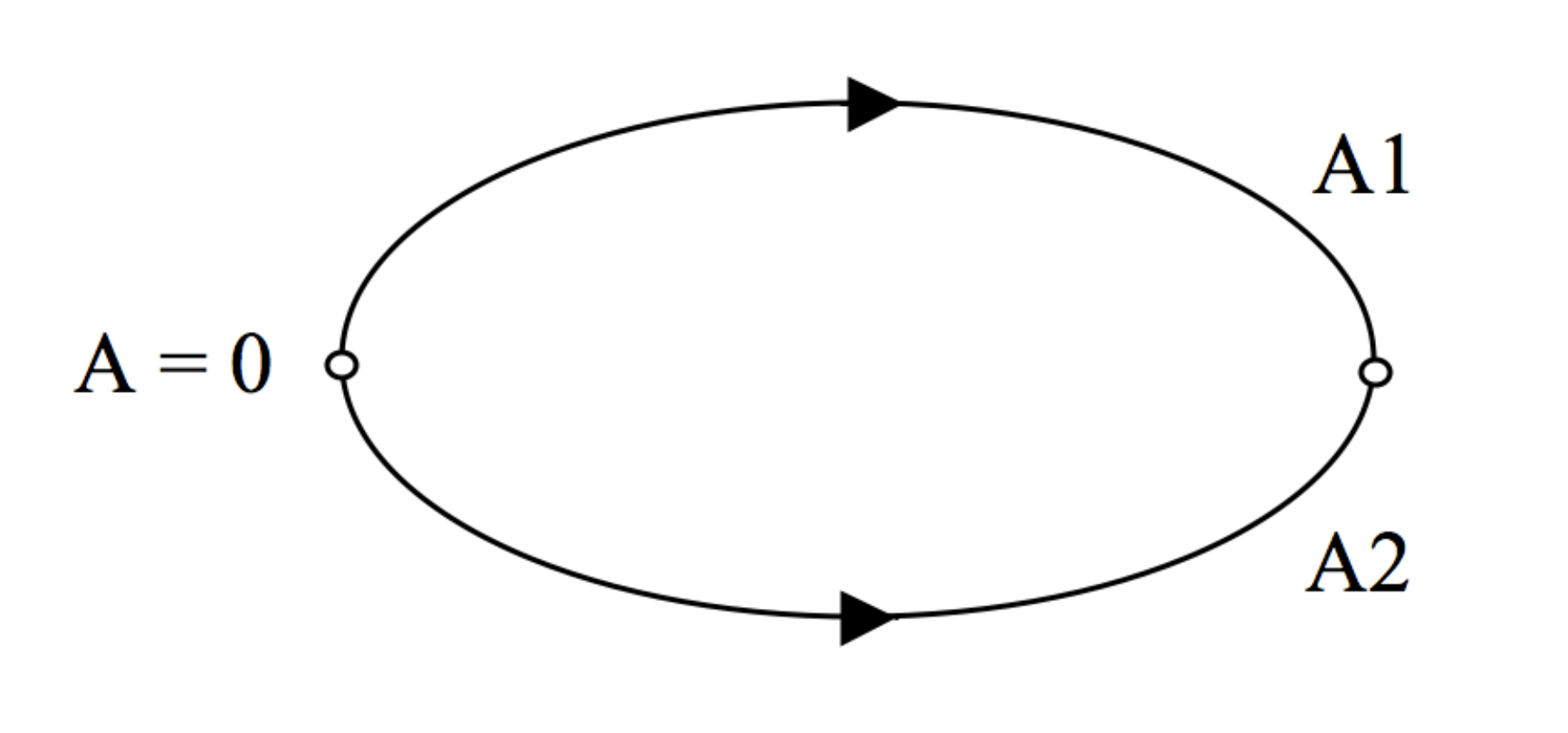}
\caption{A loop in configuration space. $A_1 (t, x)$ interpolates between the vacuum $A = 0$ and the Jackiw-Pi configuration $A_{JP} (x)$ described in the text, while $A_2 (t,x)$ interpolates between $A=0$ and a gauge transform of $A_{JP} (x)$.}
\label{figure1}
\end{figure}

\bigskip

In analogy to the Aharonov-Bohm effect, the interference pattern observed at $t = \infty$ depends on the value of theta, as the relative phase is proportional to theta times the difference in magnetic helicities,
$H(A) - H(A^\Omega)$. This difference in helicities is not quantized.

\section{Periodicity of theta dependence}

Because the difference in helicities $H(A) - H(A^\Omega)$ is not quantized, the physical effects of theta are not $2 \pi$ periodic. This is surprising because of, e.g., the following argument: in the presence of fermions we can use the anomaly relation to rotate theta onto the fermion mass matrix; because a $2 \pi$ shift in the phase of a fermion mass leaves the mass invariant, physics should therefore be $2 \pi$ periodic in theta. How is this apparent contradiction resolved?

A careful derivation \cite{GH} of the anomaly equation in the presence of general background gauge fields $A(t,x)$ (i.e., including those which approach zero as $1/r$ asymptotically) yields a correction to the usual expression:
\beq
\label{anomaly}
\partial_{\mu} j^\mu = {1 \over 16 \pi^2} F \tilde{F} ~-~ \Delta J (A)~~.
\eeq
Although we have written $\Delta J (A)$ on the right hand side of the equation, it arises due to proper regularization of the fermion current $j^\mu = \bar{\psi} \Gamma^\mu \psi$. When the background $A$ has strong asymptotic behavior the $A = 0$ subtraction used to regularize $j^\mu$ leaves a residual defined here as $\Delta J (A)$. Note that
\beq
\int d^4x ~ \Delta J ~=~ {1 \over 2} \Delta \eta (0) ~=~ {1 \over 2} \eta (0) \bigg|^{t = \infty}_{t = 0}
\eeq 
so the spacetime integral of the anomaly equation (\ref{anomaly}) yields the Atiyah-Patodi-Singer index theorem
\beq
\label{APS}
n_+ - n_- =  \int d^4x ~\partial_{\mu} j^\mu = {1 \over 16 \pi^2} \int d^4x~ F \tilde{F} -  {1 \over 2} \Delta \eta (0)~.
\eeq
Here $\Delta \eta (s)$ is the change in the ``$\eta$-invariant'' \cite{eta}
\beq
\eta (s) = \sum_{\lambda \neq 0} ~ \frac{{\rm sign} ( \lambda )} {| \lambda |^{-s} }~~,
\eeq
defined in terms of eigenvalues $\lambda$ of the Dirac operator, and $n_+ - n_-$ counts the level crossings (spectral flow) of these eigenvalues. For gauge configurations which interpolate between vacua (or which approach vacua sufficiently rapidly at early and late times), $\Delta \eta (0) = 0$, because the eigenvalues at early and late times are identical. However, for configurations such as ours, with $A \sim 1/r$ behavior, $\Delta \eta (0)$ does not vanish, because the eigenvalues of the Dirac operator do not approach their vacuum values as $t \rightarrow \infty$. The left hand side of the APS theorem (\ref{APS}) is an integer, whereas in our case both terms on the right hand side can take on fractional values. In the presence of such configurations, the effects of theta cannot simply be rotated into the fermion mass matrix.

To demonstrate the equivalence of a phase $\alpha$ in the fermion mass (assume for simplicity a single flavor) to the effect of a theta term, one makes a chiral rotation of the fermion fields, which cancels $\alpha$ in the mass term, but shifts the action by
\beq
\Delta S = \int d^4x ~ \Delta {\cal L} = \alpha \int d^4x ~\partial_{\mu} j^\mu = \alpha \, (n_+ - n_-)~~.
\eeq
This shift is periodic under $\alpha \rightarrow \alpha + 2 \pi$ because $n_+ - n_-$ is an integer. However, we see from the index theorem (\ref{APS}) or the anomaly equation (\ref{anomaly}) that the effect on the action from this chiral rotation is {\it not} equivalent to a shift in the theta term if the background field $A$ is one for which the eta invariant is non-zero:
\beq
\label{DS}
\Delta S = \alpha  \int d^4x~ \bigg[ {1 \over 16 \pi^2} F \tilde{F}   ~-~ \Delta J (A) \bigg]~~.
\eeq
In background fields (such as those with $1/r$ behavior) for which the last term in (\ref{DS}) is non-zero, a chiral rotation is not equivalent to a shift in theta -- i.e., to a shift in the coefficient of $F \tilde{F}$ only. While the physical effects due to a phase $\alpha$ in the fermion mass are invariant under $\alpha \rightarrow \alpha + 2 \pi$, the effects due to theta are not periodic, because neither the magnetic helicity nor changes in the magnetic helicity are quantized; both terms multiplying $\alpha$ in (\ref{DS}) can take on fractional or irrational values. In the presence of generic gauge backgrounds with $1/r$ behavior the theta term cannot be rotated away in favor of a phase in the fermion mass matrix.

\section{Conclusions and implications for QCD}

We described an interference effect in QED which is sensitive to the theta angle. However, this effect required fields with $A \sim 1/r$ or $B \sim 1/r^2$ behavior, which are non-perturbative and hence presumably only lead to exponentially small effects (e.g., of order $\exp (- 2 \pi / \alpha )~$). In order to be explicit we used the example of the Jackiw-Pi configurations (\ref{JP1}), but it is easy to construct other examples -- the key requirement is simply that the magnetic field have sufficiently strong asymptotic behavior. If this asymptotic behavior is allowed, then these paths occur in the functional integral and their relative phase depends on theta. Note that these configurations do not exhibit monopole magnetic charges; all magnetic field lines form closed loops.

Whether theta has physical consequences depends crucially on the boundary conditions imposed on the gauge fields. This is not surprising, as $F\tilde{F}$ is a total derivative and the theta term can be rewritten as an integral over the boundary of spacetime. However, the situation is quite different from ordinary dynamics in quantum field theory, which usually does not depend on the choice of boundary conditions once the volume is taken to infinity (special cases such as symmetry breaking excepted).

Now consider pure gauge configurations in (3+1) QCD: $A_\mu = i U^\dagger \partial_\mu U$. If the gauge function $U(|\vec{x}| \rightarrow \infty)$ approaches a constant matrix, then these configurations can be classified topologically ($U(\vec{x})$ maps $S^3 \rightarrow$ SU(3)$\,$) and have potentials $A$ that fall off faster than $1/r$. However, the condition that $U(|\vec{x}| \rightarrow \infty)$ approach a constant matrix privileges a particular position in space, and seems inconsistent with translation invariance (although the invariance is approximately good far from the boundary). Instead, we can build a translation-invariant state by superposing configurations where $U(\vec{x})$ is allowed to vary, even as $|\vec{x}| \rightarrow \infty$. For such configurations, winding numbers and topological charges can be fractional, and physics is no longer periodic in theta. As in the discussion of the previous section, the physical effects of the theta term are then {\it not} equivalent to those of a phase in the quark mass matrix, which only captures the effects of fluctuations that lead to actual level crossing: integer $n_+ - n_-$. This observation calls into question the usual calculation relating the neutron electric dipole moment and theta, which proceeds by rotating the theta angle onto the quark masses, and then uses the chiral Lagrangian.

While we do not expect the ordinary dynamics of QCD to depend on choice of boundary conditions at infinity, it seems that theta effects (such as, but not limited to, strong CP violation) are sensitive to whether one imposes integer or fractional topological charge. 

Consider the following example. Suppose the topological charge $q = (1/ 16 \pi^2) \int  F \tilde{F}$ is not quantized, and hence theta is not a compact variable. The vacuum energy density $\epsilon ( \theta )$ can be computed using the Euclidean path integral
\beq
\label{RL}
Z = \exp ( - V \epsilon (\theta) ) ~=~ \int DA~~  \exp \left( - \int d^4x ~  {1 \over 4} F^2  +   i 16 \pi^2 q \theta  \right) ~\equiv~ ~\int_{0}^{\infty}  dq  ~\mu(q)~  2 \cos ( 16 \pi^2 q \theta  )~.
\eeq
The Riemann-Lebesgue lemma implies that the integral above vanishes monotonically as $\theta \rightarrow \infty$, which implies that the energy density $\epsilon ( \theta )$ is not periodic in theta, but rather increases monotonically. It seems possible that at large enough theta phase oscillations suppress sectors with $q \neq 0$, rendering QCD nearly CP conserving. Evidence against this possibility is provided by Witten's investigation of the $\theta$-dependence of the vacuum energy in four dimensional pure gauge theory using AdS/CFT \cite{witten}, which found that periodicity under $2 \pi$ shifts in theta could be imposed even at large $N$. We note, however, that it is not clear which boundary conditions (i.e., those permitting or forbidding fractional $q$) are appropriate to the AdS/CFT duality. Historically, most authors have simply assumed that $1/r$ asymptotic gauge potentials are not allowed in gauge theory, which implies integer $q$. It is possible that AdS/CFT duality applies to integer $q$ gauge theory, but not to the more general class of boundary conditions considered here.

\newpage

\noindent {\bf Note Added} (2020): We can give a simple example of a physical quantity that is affected by $\theta$ in QED. Consider the amplitude connecting two gauge configurations $A_i$ and $A_f$:
\beq
\label{amplitude}
\langle A_f \vert U_\theta \vert A_i \rangle =  \int_{A_i}^{A_f} DA~  \exp \left( - i \int d^4x ~  {1 \over 4} F^2  +   \theta  E \cdot B  \right)  ~~~,
\eeq 
Assume that $A_f$ (specifically, the final state $B$ field) has large $r$ behavior such that the magnetic helicity is gauge dependent. Under a gauge transformation with parameter $\Omega (x)$ the change in magnetic helicity (i.e., which multiplies $\theta$ in the action above) is
\beq
\delta H = \frac{1}{2} \int d\Sigma_i ~\Omega B^i ~~.
\eeq
To obtain a physical amplitude, we must integrate over gauge copies of the potential $A$. Note all elements in (\ref{amplitude}) above are gauge invariant except the magnetic helicity term (coefficient of $\theta$) in the presence of magnetic fields that contribute to $d \Sigma_i B^i$ in the integral above. In particular, the integral over gauge functions on the spatial boundary at large $t$ is a functional integral of the form
\beq
\label{fdelta}
\int D\Omega ~ \exp \left( i \theta \int d\Sigma_i~\Omega B^i \right)   ~ \sim ~ \delta \, [ ~B^i~ ]~,
\eeq
where the argument of the functional delta function on the right hand side is the asymptotic magnetic field (at late times and large $r$). Therefore, unless $\theta = 0$, the amplitude to produce a magnetic field of this kind in the final state is exactly zero. Only when $\theta = 0$ are such amplitudes non-zero.

More generally (including in QCD and non-Abelian gauge theories): the $F \tilde{F}$ term in the action is only gauge invariant for certain choices of boundary condition on $A$, such as that it fall off faster than $1/r$. We see that other configurations (e.g., those with fractional topological charge in the non-Abelian case, or that contribute to $d \Sigma_i B^i$ in QED) do not contribute to the path integral due to the integration over gauge copies as in (\ref{fdelta}), except in the special case where $\theta = 0$.

\bigskip

\vskip 2.0cm \noindent {\bf Acknowledgments}$\,$ The author thanks Sean Carroll and Mark Wise for useful discussions. This work was
supported in part by Grant No. DE-FG02-96ER40949 from the U.S. Department of Energy and by a fellowship from the National Science Council of Taiwan. The author acknowledges the hospitality of Academia Sinica (Taipei) and Caltech while this research was conducted.


\begin{thebibliography}{99}
\bibitem{Mink} For example, the Lagrangian density of $\phi^4$ theory evaluated on a domain wall configuration yields infinite action when integrated over all of Minkowski spacetime. If ${\cal L} = \frac{1}{2} (\partial \phi)^2 - V( \phi)$, then imposing the equations of motion we obtain ${\cal L} = \frac{1}{2} \phi V'( \phi ) - V( \phi )$. For $V = ( \phi^2 - 1)^2$, we obtain ${\cal L} = ( \phi^4 - 1 )$. The spacetime region in the wall where $\phi$ interpolates between $\pm 1$ occupies infinite four-volume and contributes an infinite (negative) amount to the action.
\end{thebibliography}
\end{document}